\documentclass[prd,superscriptaddress,twocolumn,nofootinbib,showpacs,preprintnumbers,floatfix]{revtex4}

\usepackage{mathrsfs}
\usepackage{amsfonts,amssymb,amsmath}
\usepackage{graphicx,epsfig}
\usepackage{multirow}
\usepackage{verbatim}

\def\fsu5{$\cal{F}$-$SU(5)$}
\def\bfsu5{$\boldsymbol{\mathcal{F}}$-$\boldsymbol{SU(5)}$}

\def\m1half{$M_{1/2}$}
\def\m3half{$M_{3/2}$}
\def\m32{$M_{32}$}

\def\fb{${\rm fb}^{-1}$~}
\def\fbns{${\rm fb}^{-1}$}
\def\mt2{$M_{T2}$}
\def\x2{$\chi^2$}
\def\2b{$M_{T2}b$}

\def\bs0{$B_S^0 \rightarrow \mu^+ \mu^-$}

\begin{document}

\title{Primordial Synthesis: \bfsu5 SUSY Multijets, 145-150 GeV LSP,\\Proton \& Rare Decays, 125 GeV Higgs Boson, and WMAP7}

\author{Tianjun Li}

\affiliation{State Key Laboratory of Theoretical Physics and Kavli Institute for Theoretical Physics China (KITPC),
Institute of Theoretical Physics, Chinese Academy of Sciences, Beijing 100190, P. R. China}

\affiliation{George P. and Cynthia W. Mitchell Institute for Fundamental Physics and Astronomy, Texas A$\&$M University, College Station, TX 77843, USA}

\author{James A. Maxin}

\affiliation{George P. and Cynthia W. Mitchell Institute for Fundamental Physics and Astronomy, Texas A$\&$M University, College Station, TX 77843, USA}

\author{Dimitri V. Nanopoulos}

\affiliation{George P. and Cynthia W. Mitchell Institute for Fundamental Physics and Astronomy, Texas A$\&$M University, College Station, TX 77843, USA}

\affiliation{Astroparticle Physics Group, Houston Advanced Research Center (HARC), Mitchell Campus, Woodlands, TX 77381, USA}

\affiliation{Academy of Athens, Division of Natural Sciences, 28 Panepistimiou Avenue, Athens 10679, Greece}

\author{Joel W. Walker}

\affiliation{Department of Physics, Sam Houston State University, Huntsville, TX 77341, USA}


\begin{abstract}

We examine the first ATLAS Collaboration $\sqrt{s} = 8$ TeV 5.8 \fb supersymmetry (SUSY) multijet data
observations in the context of No-Scale Flipped $SU(5)$ with extra TeV-Scale vector-like flippon
multiplets, dubbed \fsu5, finding that the recent 8 TeV collider data is statistically consistent with
our prior 7 TeV results. Furthermore, we synthesize all currently ongoing experiments searching for
beyond the Standard Model (BSM) physics with this fit to the 8 TeV data, establishing a suggestive
global coherence within a No-Scale \fsu5 high-energy framework. The SUSY mass scale consistent with all BSM data consists of the
region of the \fsu5 model space within $660 \lesssim M_{1/2} \lesssim 760$ GeV, which corresponds to
sparticle masses of $133 \lesssim M({\rm \widetilde{\chi}_1^0}) \lesssim 160$ GeV,
$725 \lesssim M({\rm \widetilde{t}_1}) \lesssim 845$ GeV, and $890 \lesssim M({\rm \widetilde{g}}) \lesssim 1025$
GeV. We suggest that the tight non-trivial correspondence between the SUSY multijets, direct and indirect searches for
dark matter, proton decay, rare-decay processes, the observed Higgs boson mass, and the measured dark matter relic
density, is strongly indicative of a deeper fundamental relationship.   We additionally
suggest a simple mechanism for enhancing the capture efficiency of \fsu5 SUSY multijets, which
results in a 93\% suppression in ATLAS reported background events, but only a 27\% decrease in Monte Carlo
simulated \fsu5 multijet events.

\end{abstract}



\preprint{ACT-13-12, MIFPA-12-37}

\maketitle


\section{Introduction}

In this work we extend the projection of Monte Carlo (MC) collider-detector simulation from the No-Scale
\fsu5 model~\cite{Li:2010ws,Li:2010mi,Li:2010uu,Li:2011dw,Li:2011hr,Maxin:2011hy,Li:2011xu,Li:2011in,
Li:2011gh,Li:2011rp,Li:2011fu,Li:2011xg,Li:2011ex,Li:2011av,Li:2011ab,Li:2012hm,Li:2012tr,Li:2012ix,Li:2012yd,Li:2012qv,Li:2012jf}
onto data collected at the Large Hadron Collider (LHC) experiment into the emerging results for the 2012
$\sqrt{s} = 8$~TeV run~\cite{ATLAS-CONF-2012-103,ATLAS-CONF-2012-109}. We undertake a broader
survey of the component sources of error, both internal to our analysis and external, that may obscure a
clear delineation of the experimentally favored \fsu5 mass scale. Our primary finding is that, while the
gaugino mass centrally indicated by the newer data is discernibly elevated relative to the lower energy
2011 run, the mutual overlap between the associated 2$\sigma$ uncertainties remains sufficient to
declare the results statistically consistent.

A perfectly direct comparison between the $\sqrt{s} = 7$~TeV~\cite{ATLAS-CONF-2012-033,ATLAS-CONF-2012-037} and available
$\sqrt{s} = 8$~TeV~\cite{ATLAS-CONF-2012-103,ATLAS-CONF-2012-109} search strategies is complicated in one
studied example~\cite{ATLAS-CONF-2012-109} by modifications to the channel segregation and
selection cuts that have presumably been employed for optimization of the expected cross section limits
for certain simplified Supersymmetry (SUSY) models. In another case~\cite{ATLAS-CONF-2012-103}, we presently suggest
an alternative selection cut threshold derived out of the finely grained bin-wise results supplied by the
ATLAS collaboration that has the dual benefit of i) improving the statistical correlation between the
published 2011 and 2012 results, and ii) substantially further taxing the strength of the Standard Model
(SM) background competition while assessing only a comparatively modest duty against the expected \fsu5
signal. Simultaneously, we remain mindful of alternative factors that may yet
play a role in the ultimate interpretation of contemporary circumstances, such as the difficulty in
establishing supersymmetric cross sections at the Next to Leading Order (NLO)~\cite{Badger:2012pf},
and the possibility of an understandably cautious default institutional stance in the estimation
of event backgrounds~\cite{Nachman:2012zf}. Moreover, although the relevant SUSY cross sections
associated with collider operation at $\sqrt{s} = 8$~TeV may be a few-fold enhanced relative to those at
$\sqrt{s} = 7$~TeV, the luminosities integrated in either case remain for now comparable, and it
similarly remains somewhat unclear whether the onus of reconciliation between these observations will
come to rest more so with an upward statistical fluctuation in the former season of data collection or a
downward fluctuation in the latter.

We emphasize that our sympathy for the No-Scale \fsu5 model derives not out of any single collider based
result, but rather from an aggregation of successful comparisons to data on the Cold Dark Matter
(CDM) relic density, the mass of the lightest Charge conjugation-Parity (CP) even Higgs
field, post-SM contributions to rare processes, Astro-physical line-sources in the gamma-ray
spectrum, and proton decay, coupled with deep structural motivations that include basic string- and
F-theoretic model building consistency, the reductionism and cosmological relevance of No-Scale
Supergravity, and the full range of phenomenological advantages associated with the Flipped $SU(5)$ Grand Unified
Theory (GUT).  The specificity afforded by this tightly defined global framework is both the mechanism by
which detailed comparison against the collider results is facilitated, and also an agent of intuition and
context for the rationalization and interpretation of those results. This is not to our perception wishful,
but rather a consciously guided logic corresponding to the net accumulation of prior probabilities in the
Bayesian sense. It is indeed the inclusive primordial synthesis of diverse experimentation and well motivated theory.

\section{The \fsu5 Model}

Supersymmetry naturally solves
the gauge hierarchy problem in the SM, and suggests (given $R$ parity conservation)
the lightest supersymmetric particle (LSP) as a suitable cold dark matter candidate.
However, since we do not see mass degeneracy of the superpartners,
SUSY must be broken around the TeV scale. In GUTs with
gravity mediated supersymmetry breaking, called
the supergravity models,
we can fully characterize the supersymmetry breaking
soft terms by four universal parameters
(gaugino mass $M_{1/2}$, scalar mass $M_0$, trilinear soft term $A$, and
the low energy ratio of Higgs vacuum expectation values (VEVs) $\tan\beta$),
plus the sign of the Higgs bilinear mass term $\mu$.

No-Scale Supergravity was proposed~\cite{Cremmer:1983bf}
to address the cosmological flatness problem,
as the subset of supergravity models
which satisfy the following three constraints:
i) the vacuum energy vanishes automatically due to the suitable
 K\"ahler potential, ii) at the minimum of the scalar
potential there exist flat directions that leave the
gravitino mass $M_{3/2}$ undetermined, iii) the quantity
${\rm Str} {\cal M}^2$ is zero at the minimum. If the third condition
were not true, large one-loop corrections would force $M_{3/2}$ to be
either identically zero or of the Planck scale. A simple K\"ahler potential that 
satisfies the first two conditions is~\cite{Ellis:1984bm,Cremmer:1983bf}
\begin{eqnarray} 
K &=& -3 {\rm ln}( T+\overline{T}-\sum_i \overline{\Phi}_i
\Phi_i)~,~
\label{NS-Kahler}
\end{eqnarray}
where $T$ is a modulus field and $\Phi_i$ are matter fields.
The third condition is model dependent and can always be satisfied in
principle~\cite{Ferrara:1994kg}.
For the simple K\"ahler potential in Eq.~(\ref{NS-Kahler})
we automatically obtain the No-Scale boundary condition
$M_0=A=B_{\mu}=0$ while $M_{1/2}$ is allowed,
and indeed required for SUSY breaking.
Because the minimum of the electroweak (EW) Higgs potential
$(V_{EW})_{min}$ depends on $M_{3/2}$,  the gravitino mass is 
determined by the equation $d(V_{EW})_{min}/dM_{3/2}=0$.
Thus, the supersymmetry breaking scale is determined
dynamically. No-scale supergravity can be
realized in the compactification of the weakly coupled
heterotic string theory~\cite{Witten:1985xb} and the compactification of
M-theory on $S^1/Z_2$ at the leading order~\cite{Li:1997sk}.

In order to achieve true string-scale gauge coupling unification
while avoiding the Landau pole problem,
we supplement the the standard ${\cal F}$-lipped $SU(5)\times U(1)_X$~\cite{Nanopoulos:2002qk,Barr:1981qv,Derendinger:1983aj,Antoniadis:1987dx}
SUSY field content with the following TeV-scale vector-like multiplets (flippons)~\cite{Jiang:2006hf}
\begin{eqnarray}
\hspace{-.3in}
& \left( {XF}_{\mathbf{(10,1)}} \equiv (XQ,XD^c,XN^c),~{\overline{XF}}_{\mathbf{({\overline{10}},-1)}} \right)\, ,&
\nonumber \\
\hspace{-.3in}
& \left( {Xl}_{\mathbf{(1, -5)}},~{\overline{Xl}}_{\mathbf{(1, 5)}}\equiv XE^c \right)\, ,&
\label{z1z2}
\end{eqnarray}
where $XQ$, $XD^c$, $XE^c$, $XN^c$ have the same quantum numbers as the
quark doublet, the right-handed down-type quark, charged lepton, and
neutrino, respectively.
Such kind of models can be realized in ${\cal F}$-ree ${\cal F}$-ermionic string
constructions~\cite{Lopez:1992kg},
and ${\cal F}$-theory model building. Thus, they have been 
dubbed ${\cal F}$-$SU(5)$~\cite{Jiang:2009zza}.

\section{The Role of Model-Specific Predictions}

Significant particle physics searches of the past few decades had the benefit of a well developed
Standard Model framework, leading to reliable theoretically based predictions and expectations.
By contrast, the view available for steering the ongoing SUSY search is rather more clouded,
with the symmetry breaking mechanism, the boundary conditions and level of universality
interrelating high-energy model parameters, and even the reality of low-energy SUSY itself
still unknown.  The LHC SUSY search effort is at the present, relatively speaking, in its very early stages. While the
collaborations must understandably prefer a model-agnostic point of view, and insist on a much larger
accumulation of data before a possible presence of beyond the Standard Model (BSM) physics is acknowledged,
our role as a small theory group is distinct, yet complementary.
Crucially, we have the unique freedom to focus not on the general, but rather the specific.  Nature, after all,
is the physical embodiment of but one single framework, possessed of features that may be cloaked to the parallel
consideration of all conceivable morphologies.  Given the rapid encroachment upon the leading BSM candidates' model
spaces that has been observed during the first phase of LHC operation~\cite{Buchmueller:2012mc}, the likelihood 
that alternatively viable conceptions such as No-Scale \fsu5 may now drive the discussion has become
sharply heightened.  In particular, the specificity of such a context may both facilitate
the global interpretation of experimental results and suggest preferred tactics for future searches. 

A central question facing any attempt to compare a specific SUSY model construction against the emerging LHC data
is that of how to correctly interpret mass limits published by the collaborations that are derived under the
assumption of a highly simplified particle content or an alternatively conjectured theoretical SUSY framework,
typically of the CMSSM or mSUGRA variety.  Such examples often decouple all SUSY fields other than the neutralino and gluino octet,
or otherwise impose strict mass degeneracies, and may unrealistically force key branching ratios to unity.  Our reading of limits
of this type is that their merit is largely as a standardized metric for quantifying the overall ``strength'' of the search
relative to prior surveys, whereas the applicability to any given physical scenario is somewhat less direct, and highly model dependent.
In particular, it is well known that breaking the squark mass degeneracy into a strongly differentiated hierarchy may
result in the substantial production of low transverse momentum jets that can evade currently favored thresholds for reconstruction.
As such, we caution against an overly literal application of the mass limits established via these surveys to the \fsu5 context.
Moreover, we suggest that studies such as our own may fill a vital gap of information by providing
independent data-driven demarcation of mass limits specifically applicable to certain realistic models,
inclusive of all applicable sources of uncertainty. 

The adoption of an exclusively specified analysis framework also affords some additional measure of resolution
in the attempt to associate collider results from individual search channels with the signature
production modes of the model under study.  Apparently isolated details that might be overlooked in a
model-independent approach may be granted additional significance within the context of a unifying model
that allows for their mutual correlation.  In conjunction, a decoupling of the corporate inertia associated
with the need to assimilate multiple incompatible perspectives allows for a more lithe response to
the volatile environment of a rapidly accumulating data store.
Having said that, our job is not to replace, nor even to reproduce, the ATLAS experimental SUSY search
analyses of Refs.~\cite{ATLAS-CONF-2012-103,ATLAS-CONF-2012-109}.  Rather, our focus is simply to
assemble clues relevant to determination of the fate of one particular highly realistic model that 
might escape comment within a more diluted scope.  Whereas ATLAS and CMS might bet on black or on red, we
shall bet on green.  The odds are certainly longer, but the payout may be commensurately greater.

\section{8 TeV SUSY Multijets}

The No-Scale \fsu5 model leverages the $\beta$-function coefficient $b_i$ modifications induced by inclusion of the
vector-like flippon multiplets to flatten the $SU(3)$ Renormalization Group Equation (RGE) running ($b_3 = 0$)~\cite{Li:2010ws},
generating a characteristic mass texture of $M(\widetilde{t}_1) < M(\widetilde{g}) < M(\widetilde{q})$, featuring a light stop and
gluino that are lighter than all other squarks. This distinctive hierarchy is highly stable across the full
viable parameter space, and is rescaled {\it en masse} according to action of the isolated dimensionful
parameter $M_{1/2}$.  We are not aware of any CMSSM/mSUGRA constructions in which this mass ordering
is precisely replicated.  The small light stop and gluino mass splitting is generated by the same strongness of the
Higgs to top quark couplings that provide the essential lifting of the MSSM Higgs boson mass. A direct consequence
of this spectrum is a distinctive event topology beginning with the pair-production of squarks
$\widetilde{q}$ and/or gluinos $\widetilde{g}$ in the initial hard scattering process, where each
squark will most likely produce gluinos via $\widetilde{q} \to q \widetilde{g}$. Gluinos will then
likely decay either through light stops by
$\widetilde{g} \to \widetilde{t}_1 \overline{t} \to t \overline{t} \widetilde{\chi}_1^0$,
or through $\widetilde{g} \to q \overline{q} \widetilde{\chi}_1^0$, with each gluino producing 2--6 jets.
Therefore, the gluino-mediated stop channel diagram, which is most probable in \fsu5, can produce up to 12
jets from a single pair-production gluino event~\cite{Maxin:2011hy,Li:2011hr}.

The unique rescaling of the No-Scale \fsu5 SUSY particle hierarchy with respect to variation of $M_{1/2}$
implies a progressive widening of the mass-gap separating the gluino and light stop that yields
a very interesting threshold transition.  Specifically, the production of light stops in the previously
described gluino-mediated channel abruptly shifts from off-shell to on-shell at $M_{1/2} = 729$ GeV, where
$M(\widetilde{g}) - M(\widetilde{t}_1) = M(t)$.  This effect will produce non-ignorable consequences for 
certain collider signatures, driven by the fact that
$\widetilde{g} \to \widetilde{t}_1 \overline{t}$ occurs at a 100\% branching fraction for
$M_{1/2} > 729$ GeV, effectively suppressing $\widetilde{g} \to q \overline{q} \widetilde{\chi}_1^0$.
We shall examine this transition with care here and account for its effect in our results.

We have previously extensively studied~\cite{Li:2011av,Li:2012hm,Li:2012tr,Li:2012ix} the ATLAS and CMS 1, 2, and 5 \fb LHC data
observations in the context of \fsu5.  Small, but curious, excesses beyond the Standard Model
expectations were observed in the initial 1--2 \fb of multijet data at
$\sqrt{s} = 7$~TeV~\cite{PAS-SUS-11-003,Aad:2011qa,Aad:2011ib}, which we demonstrated could 
be neatly and globally explained as supersymmetry production~\cite{Li:2011fu,Li:2011av,Li:2012hm}
within the No-Scale \fsu5 model.  Assuming a purely authentic signal, we projected the anticipated
growth forward to 5 \fb at 7 TeV.  Upon subsequent release of the corresponding 5 \fb experimental
results, we found these projections to be well substantiated in the data superset~\cite{Li:2012tr,Li:2012ix},
with a consistently isolated SUSY mass scale and a commensurate signal intensification.
We interpreted this finding as an amplification of the likelihood that the small observed excesses
were indeed largely attributable to fundamental physics, and not random background statistical fluctuation.

Completion of the first $\sqrt{s} = 8$ TeV SUSY multijet searches by
ATLAS~\cite{ATLAS-CONF-2012-103,ATLAS-CONF-2012-109} compels extension of
the comparative \fsu5 analysis undertaken at 7 TeV~\cite{Li:2011av,Li:2012hm,Li:2012tr,Li:2012ix}
into the elevated beam collision energy context.  Our effort is somewhat complicated at the outset by
irreversible alterations to the 4-jet, 5-jet, and 6-jet SUSY search strategies of
Ref.~\cite{ATLAS-CONF-2012-109} from the previous 7 TeV partitions~\cite{ATLAS-CONF-2012-033}.
In the case of those SUSY searches adhering to a $\ge7$, $\ge8$, and $\ge9$ cut on jets~\cite{ATLAS-CONF-2012-103}, we
observe that a more accommodative cut for discovery of \fsu5 can be implemented by raising the ATLAS imposed
${\rm E_T^{Miss} / \sqrt{H_T}} > 4 ~{\rm GeV}^{1/2}$ requirement to a harder $> 8 ~{\rm GeV}^{1/2}$.
Averaged across all published search channels at both 7 and 8 TeV, this escalation further reduces the
already severely truncated SM background component extracted from the graphical bin-by-bin ATLAS analysis
by a full 93\%, while preserving a strong majority of critical \fsu5 SUSY events, which experience
only a 27\% reduction in our Monte Carlo simulation.  Such drastic increases in efficiency
being rather hard to come by, this simple modification is one that we heartily recommend for
further investigation to our experimental colleagues at the LHC collaborations.  Moreover, we find that the residually
observed signal proportion scales much more linearly between the 7 and 8 TeV results with this modification
than without.  In fact, the suggested lifting of the minimum boundary on the observable ratio
${\rm E_T^{Miss} / \sqrt{H_T}}$ has existing precedent at ATLAS in Ref.~\cite{ATLAS-CONF-2012-073}, where
a threshold of $11~{\rm GeV}^{1/2}$ was imposed.  Curiously, a simple visual inspection of Figure (5) in
Ref.~\cite{ATLAS-CONF-2012-073} also clearly indicates that a hard cut of
${\rm E_T^{Miss} / \sqrt{H_T}} > 11 ~{\rm GeV}^{1/2}$ implies all surviving signal excesses
in this SUSY light stop search must emanate from events with 9--10 jets.  Event production in
this ultra-high jet multiplicity regime is the leading collider signature of \fsu5, wholly in
concordance with our originally suggested search prescription~\cite{Maxin:2011hy,Li:2011hr},
providing ample justification for a more harsh and efficient cut on ${\rm E_T^{Miss} / \sqrt{H_T}}$,
as we shall apply in our work here.

To most effectively portray \fsu5 against the circumstance of the rather light data accumulation to date,
we take a cue from the recent LHC Higgs boson search, where of the five expected production channels, only the
$\gamma \gamma$ and $ZZ$ channels were initially productive in 2011, and thus utilized for compelling
evidence of Higgs boson production. In tandem, the $\tau \tau$, $WW$, and $b \overline{b}$ channels
either displayed under-production or no evidence at all, though these unproductive channels did not sway
the convincing argument in December 2011 that indeed the Higgs boson was on the verge of eclipsing a significant
observational threshold.  We believe that an extension of this search tactic is suitable to our
current purpose, and thus elect to partition the available data channels into two categories: i) inclusive,
representing those productive searches demonstrating some activity over and above the expectations for
the SM background, and ii) exclusive, denoting those searches embodying a null set of events beyond
expectations.  Within the inclusive data set, we search for correlations in the form of a consistently
isolated \fsu5 mass scale associable with each observed excess.  For the exclusive data sets, we
verify overlap of the suggested lower bound on $M_{1/2}$ with the value favored by the former best fit.
We utilize the signal significance metric ${\rm S = 2 \times (\sqrt{N_S + N_B} - \sqrt{N_B})}$ to enact
the described partition, selecting those multijet searches from Refs.~\cite{ATLAS-CONF-2012-103,ATLAS-CONF-2012-109}
that exceed a specified minimum level of significance.  For our study here, we elect a lower limit of
${\rm S = 2 \times (\sqrt{N_S + N_B} - \sqrt{N_B})} \ge 1$.

We find that only five of the seventeen multijet searches satisfy
the ${\rm S = 2 \times (\sqrt{N_S + N_B} - \sqrt{N_B})} \ge 1$ condition: 7j80, 8j55, and 8j80 of
Ref.~\cite{ATLAS-CONF-2012-103}, in conjunction with SRE Loose and SRE Medium of
Ref.~\cite{ATLAS-CONF-2012-109}. The 7j80 requires a minimum of 7 jets and jet ${\rm p_T > 80}$ GeV, while
8j55 applies a cut at $\ge8$ jets and ${\rm p_T > 55}$ GeV, with 8j80 also requiring at least 8 jets with
${\rm p_T > 80}$ GeV. The SRE Loose and SRE Tight both retain only those events with 6 jets and ${\rm p_T > 60}$ GeV,
though the Loose prescription sets a cut on effective mass, defined to be the scalar sum of the transverse
momenta of the leading N jets together with ${\rm E_T^{Miss}}$, at ${\rm M_{eff} > 1}$ TeV and
${\rm E_T^{Miss} / M_{eff} > 0.3}$, whereas the Medium case applies the cut at ${\rm M_{eff} > 1.3}$ TeV and
${\rm E_T^{Miss} / M_{eff} > 0.25}$.  Although we would like to see a somewhat lower threshold on
the reconstructed jet transverse momentum, and have advocated a value as low as ${\rm p_T \gtrsim 20}$ GeV
in Refs.~\cite{Maxin:2011hy,Li:2011hr}, we acknowledge the technical hurdles which make such an extension
difficult.  However, this practical limitation does leave a large region of the SUSY parameter
space unexplored, to which the commonly promoted simplified SUSY mass exclusion diagrams may be quite
sensitive.  We tender for comparison the case of the ATLAS light stop search in Ref.~\cite{ATLAS-CONF-2012-073},
which seemingly successfully realized a soft ${\rm p_T}$ cut on jets of ${\rm p_T > 25}$ GeV, obtaining
in the process crucial information on excess production in events with 9--10 jets.

The salvaging of these five strategies from the full ATLAS SUSY multijet 8 TeV search campaign affords us
with five degrees of freedom (DOF) with which to construct a multi-axis $\chi^2$ fitting procedure.
However, these five DOF are not all statistically independent, supplying us with an effective DOF count
which is somewhat less than five.  While we presently forgo a detailed weighting and recombination procedure,
we have explicitly verified that a reduction to three effective DOF's only marginally collapses the width
of the $\chi^2$ error margins, corresponding specifically to an elevation of the 2$\sigma$ lower SUSY mass limit
intersection by about $10$~GeV.  Additionally, we have generically quantified the correlated overlap existing
between Monte Carlo events surviving each considered search strategy, according to the prescription
${\rm N_{Common} / \sqrt{N_A \times N_B}}$.  This statistic is reported in Table~\ref{tab:correlations} for each
pairing of channels A and B, where ${\rm N_{Common}}$ is the count of overlapping events, and ${\rm N_A}$, ${\rm N_B}$
are the counts of events passing each selection individually.  We observe a modest correlation averaging about
$\sim50$\% amongst only the $\ge7$-jet, $\ge8$-jet, and $\ge9$-jet multijet searches internal to
Ref.~\cite{ATLAS-CONF-2012-103}, compared to a very small overlap of the $\ge7$-jet, $\ge8$-jet, and
$\ge9$-jet searches with the 6-jet searches of Ref.~\cite{ATLAS-CONF-2012-109}.  Likewise, the SRE
Loose and SRE Medium exhibit a very high correlation, in line with expectations.

\begin{table}[htp]
	\centering
	\caption{Values of Correlation Function measuring statistical independence amongst all SUSY searches
satisfying the condition ${\rm 2 \times (\sqrt{N_S + N_B} - \sqrt{N_B})} \ge 1$. A higher value indicates a
higher overlap of events between the two searches, where the Correlation Function can assume percentage
values from 0 to 100.}
		\begin{tabular}{|c||c|c|c|c|} \hline
			& 	8j55	&	8j80	&	SREL	&	SREM 		\\ \hline \hline
$	{\rm 7j80}	$&$	58.7	$&$	55.5	$&$	15.0	$&$	24.5	$	\\	\hline
$	{\rm 8j55}	$&$	-	$&$	43.1	$&$	23.2	$&$	32.8	$	\\	\hline
$	{\rm 8j80}	$&$	-	$&$	-	$&$	7.1	$&$	12.7	$	\\	\hline
$	{\rm SREL}	$&$	-	$&$	-	$&$-		$&$	68.6	$	\\	\hline
		\end{tabular}
		\label{tab:correlations}
\end{table}

The No-Scale \fsu5 experimentally viable parameter space features a distinctive abrupt transition from
off-shell light stop production to on-shell. We have noted this trait in prior
work~\cite{Li:2012hm,Li:2012tr}, though we had not previously explicitly taken this interesting
aspect into account in our SUSY mass fittings. We shall ascertain the entire consequence of this sharp
transition in this work, the effect of which we determine to be non-negligible. We further attempt to
integrate a more realistic compilation of all the experimental and theoretical uncertainties into our
analysis, a comprehensive treatment appended to our more abridged policy toward accumulated
uncertainties in past work.  The mentioned uncertainty may arise from many sources, the most obvious being
the Poisson statistical fluctuations of the reported event counts, the official collaboration estimates of the SM background
error, and imperfections in our own Monte Carlo event generation, detector simulation, and event
selection processing phase.  For instance, the typical margin of error on the ${\tt PGS4}$~\cite{PGS4} detector
simulation we utilize is quoted at 20\%, with the caveat that circumstances may
sometimes conspire to produce a factor as large as two.  There are also more subtle potential sources, both internal and external,
of systematic uncertainty such as i) limitations on computational accuracy of the flippon-modified \fsu5
SUSY spectrum and Higgs boson mass, and 2) the possibility that the NLO QCD corrections to multijet production
cross sections could be somewhat smaller than, perhaps even only 60\% of, the Collaboration estimates~\cite{Badger:2012pf},
as applied to the simulation of both the supersymmetry and SM background event production rate.  This
latter effect would result in an overestimation of background levels and over-production of SUSY events in the
popular Monte-Carlo codes, potentially masking the observation of legitimate experimental
SUSY signatures, while simultaneously falsely escalating the SUSY mass scale required for sufficient
event suppression.  This could also account for the suggestion of an apparent bias toward large
p-values in the global analysis of early LHC results~\cite{Nachman:2012zf}.  Our approach for 
addressing these difficulties in the present work is two-fold.  Firstly, we undertake the described 
quantification of the chi-square deviation separating the experimental results from the \fsu5 collider-detector
simulation for a continuous string of $M_{1/2}$ values spanning the otherwise viable model space.
Secondly, we repeat this analysis while enhancing or suppressing the nominal Monte Carlo event count
by a selected overall rescaling factor, adopting the outer convolution of the mass boundaries
as limits on the model.

\begin{figure}[htp]
        \centering
        \includegraphics[width=0.48\textwidth]{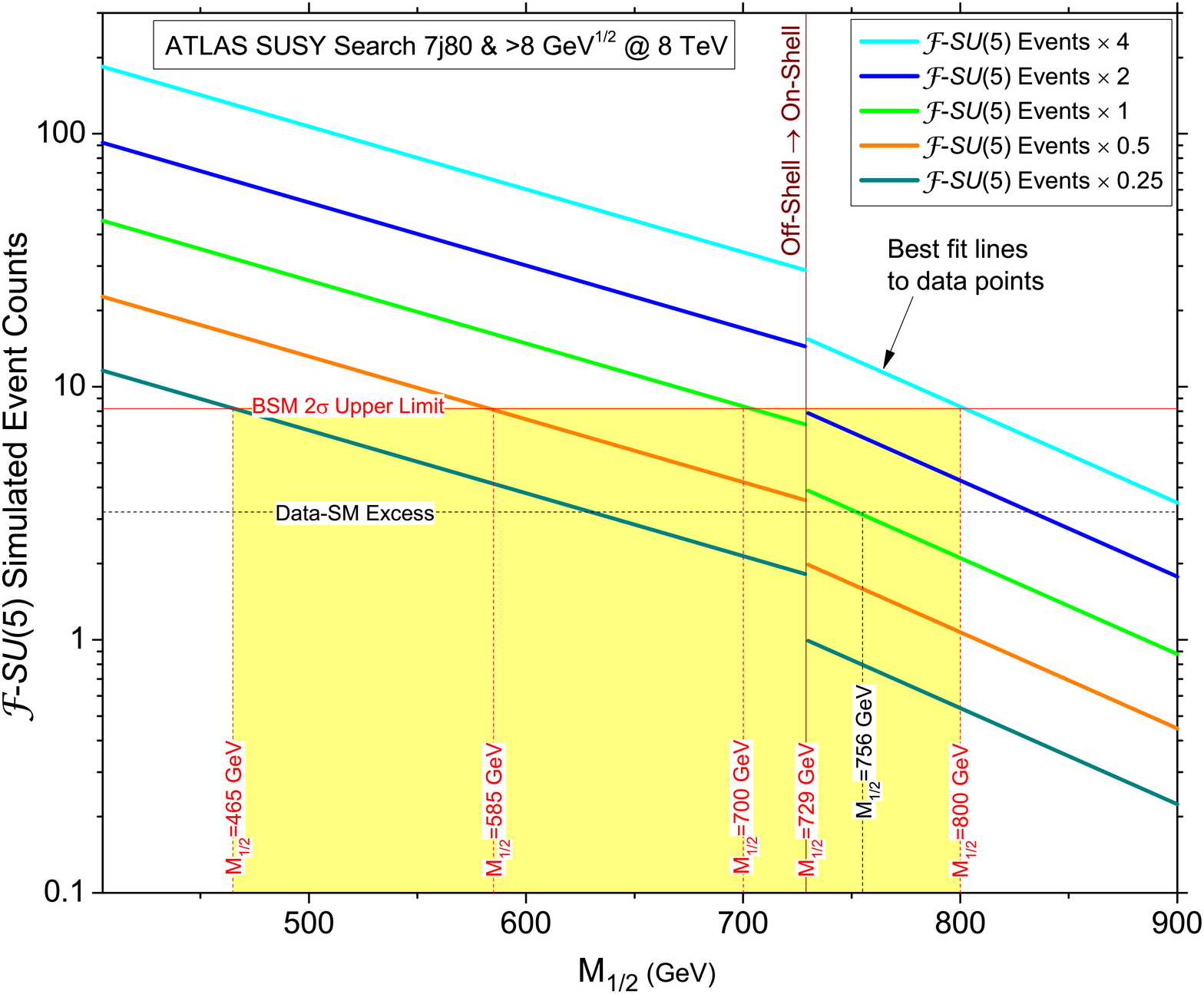}
        \caption{Measure of uncertainty in the complete sequence of SUSY mass spectrum calculations through the detector
simulations in No-Scale \fsu5 for the 7j80 SUSY multijets search using the optimized cut of
${\rm E_T^{Miss}} / \sqrt{\rm H_T} > 8 ~{\rm GeV}^{1/2}$. We increase and decrease the nominal number of \fsu5 events
surviving all the ATLAS prescribed cuts by factors of two and four to ascertain the effect of accumulated uncertainties
on the best fit SUSY mass scale. We show the 2$\sigma$ BSM upper limit on the number of collider events for each
multiple of the nominal number of events. The yellow region illustrates the total uncertainty on the 2$\sigma$
BSM upper limit when taking into account all experimental and theoretical uncertainties.}
        \label{fig:upperlimit}
\end{figure}

The consequence of compounded uncertainties on the SUSY mass spectrum in our \fsu5 computations is
depicted in Figure (\ref{fig:upperlimit}) for the selected search case of 7j80.  Shown are the nominal number of
surviving events after all data cuts, and also the relative increase and decrease by
factors of two and four.  Rescaling by a factor as large as four might be interpreted as a worst-case scenario,
although we deem the more modest factor of two to be quite reasonably necessary. The data lines in Figure
(\ref{fig:upperlimit}) represent linear fittings to a specified set of benchmarks, the
characteristics of which shall be elaborated upon in the next section. Graphically displayed is the
off-shell to on-shell transition, an aspect alone that reduces the observed cross section for
$M_{1/2} > 729$ GeV by nearly a factor of two for the case of 7j80. For 7j80, we see that the best fit to the nominal number
of excess events above the background expectations reported by ATLAS is $M_{1/2} = 756$ GeV, with a BSM
2$\sigma$ upper limit falling at $M_{1/2} \ge 700$ GeV. On the other hand, adjustments upward and downward
by the fixed multiples of two and four translate the BSM 2$\sigma$ upper limit to $M_{1/2} \ge
700^{+100}_{-235}$ GeV, represented by the yellow region in Figure (\ref{fig:upperlimit}),
permitting the majority of the \fsu5 viable parameter space to remain experimentally plausible. This
corresponds to sparticle masses of $M({\widetilde{\chi}_1^0}) \ge 142^{+25}_{-55}$ GeV,
$M({\widetilde{t}_1}) \ge 775^{+114}_{-281}$ GeV, and $M({\widetilde{g}}) \ge 944^{+134}_{-309}$
GeV (see bottom scales in the ``Primordial Synthesis'' Figure~(\ref{fig:primordial}) for $M_{1/2}$ to
gluino and light stop mass conversions).

The completion of a multi-axis $\chi^2$ analysis is facilitated by execution on twenty-one benchmark samples an in-depth Monte Carlo collider-detector simulation of all 2-body SUSY processes based on the {\tt MadGraph}~\cite{Stelzer:1994ta,MGME} program suite, including the {\tt MadEvent}~\cite{Alwall:2007st}, {\tt PYTHIA}~\cite{Sjostrand:2006za} and {\tt PGS4} chain. The SUSY particle masses are calculated with {\tt MicrOMEGAs 2.4}~\cite{Belanger:2010gh}, applying a proprietary modification of the {\tt SuSpect 2.34}~\cite{Djouadi:2002ze} codebase to run the flippon-enhanced RGEs. We implement a modified version of the default ATLAS detector specification card provided with {\tt PGS4} that calls an anti-kt jet clustering algorithm, indicating an angular scale parameter of $\Delta R = 0.4$.  The resultant event files are filtered according to a precise replication of the selection cuts specified by the ATLAS Collaboration in Refs.~\cite{ATLAS-CONF-2012-103,ATLAS-CONF-2012-109}, employing the script {\tt CutLHCO 2.0}~\cite{Walker:2012vf} to implement the post-processing cuts.

\begin{figure*}[htp]
        \centering
        \includegraphics[width=0.48\textwidth]{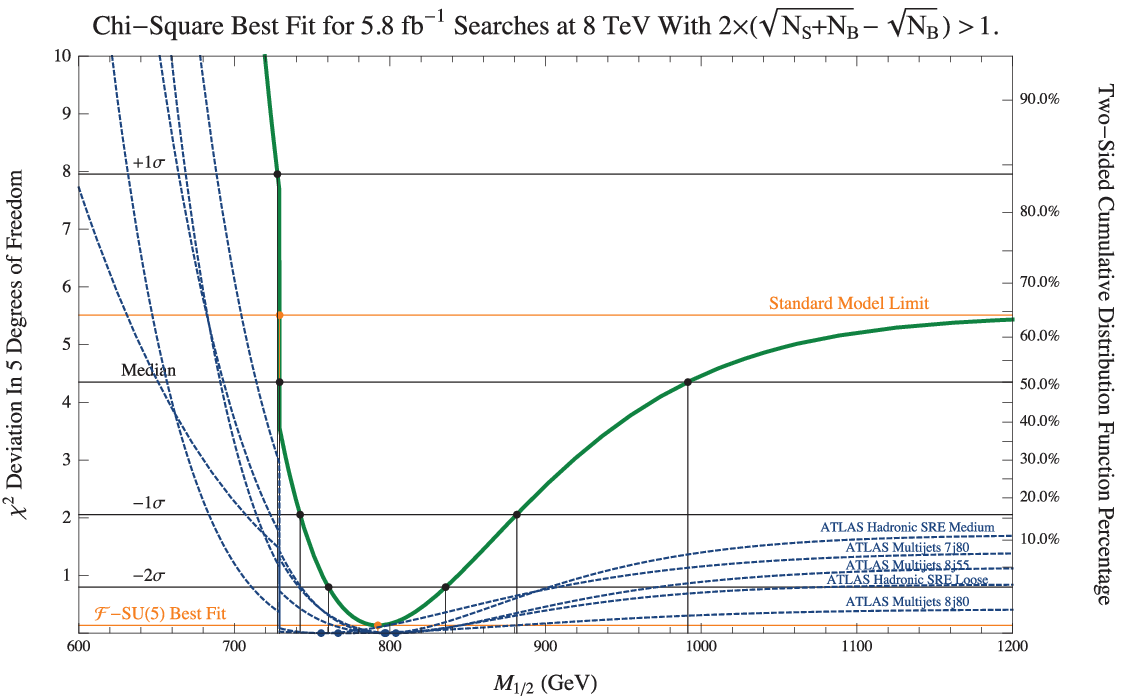}
        \includegraphics[width=0.48\textwidth]{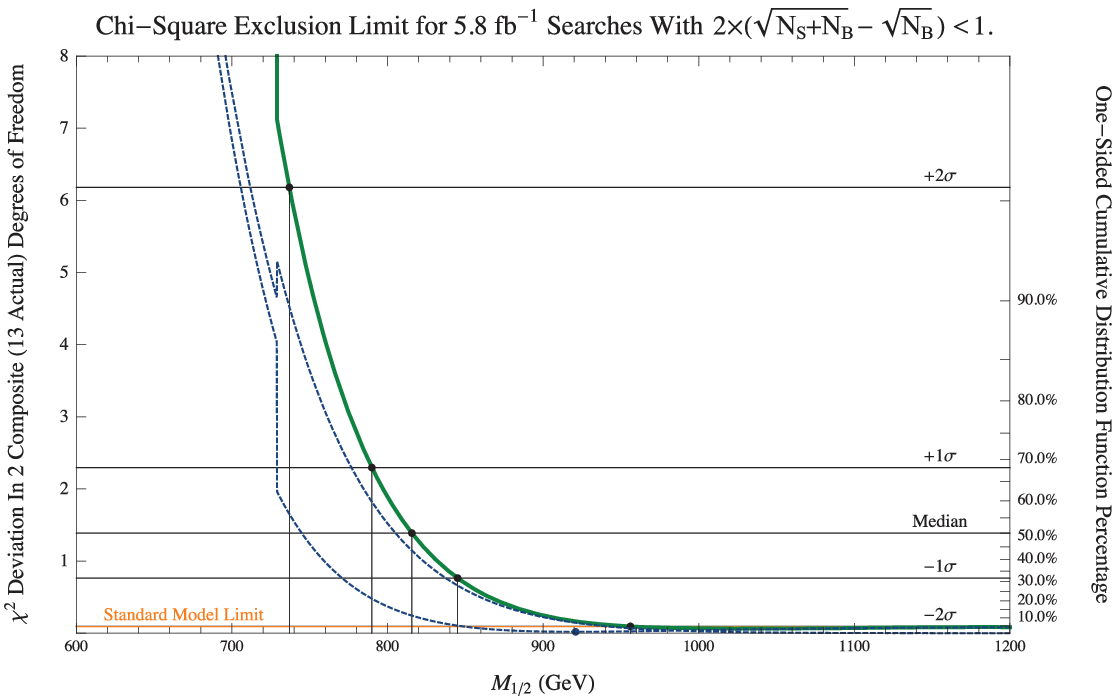}
        \caption{We depict the $\chi^2$ analyses of the 8 TeV ATLAS 5.8 \fb 7j80, 8j55, 8j80, SRE Loose, and SRE Medium Multijet search
strategies from Refs.~\cite{ATLAS-CONF-2012-103,ATLAS-CONF-2012-109} that exhibit a signal
significance of ${\rm 2 \times (\sqrt{N_S + N_B} - \sqrt{N_B})} \ge 1$ (left-hand pane), and the corresponding
null searches of Refs.~\cite{ATLAS-CONF-2012-103,ATLAS-CONF-2012-109} with
${\rm 2 \times (\sqrt{N_S + N_B} - \sqrt{N_B})} < 1$ (right-hand pane). The thin dotted blue lines correspond to the individual
$\chi^2$ curves for each event selection for only the case of the nominal number of \fsu5 events, which are
summed into the thick green cumulative multi-axis $\chi^2$ curves. The large discontinuity shown
represents the transition from off-shell to on-shell light stop production at $M_{1/2} = 729$ GeV.}
        \label{fig:chi_square}
\end{figure*}

Our $\chi^2$ analysis of the initial $\sqrt{s} = 8$~TeV ATLAS multijet searches~\cite{ATLAS-CONF-2012-103,ATLAS-CONF-2012-109}
is depicted in the two panels of Figure~(\ref{fig:chi_square}), first
for the five inclusive search strategies with signal significance ${\rm 2 \times (\sqrt{N_S + N_B} - \sqrt{N_B})} \ge 1$
and next for the thirteen channels showing no corresponding visible excess above the SM background.
The Cumulative Distribution Function (CDF) percentage labeled on the right-hand axis of each plot is a statistical
tool that establishes the fraction of trials (for a fixed number of statistically independent variables) where
Gaussian fluctuation of each of variable will yield a net deviation from the null hypothesis that is not larger
than the corresponding $\chi^2$ value referenced on the left-hand axis.  The distinction between a one-sided
and two-sided limit is made with regards to the placement of the $0,1,2$-$\sigma$ bounds relative to the numerical
value of the CDF.  A two-sided limit, as adopted in the initial panel for the $\chi^2$ best-fit minimization against
$M_{1/2}$ using searches with visible excesses, is appropriate when meaningful deviations may be anticipated in either direction away
from the median.  Particularly small $\chi^2$ values imply a better fit to the data by the assumed
signal (at a certain confidence level) than what could be attributed to random fluctuations around the SM, while excessively
large $\chi^2$ values disfavor a given $M_{1/2}$ relative to the SM-only null hypothesis.  In this case, the usual
$1$ and $2$-$\sigma$ $68\%$ and $95\%$ consistency integrations enclose areas symmetrically distributed about the mean,
such that centrally inclusive boundary lines are drawn at the CDF percentages $2.2\%$, $15.9\%$, $84.1\%$, and $97.7\%$.
It should be noted that this method implicitly assumes the displayed channels to be statistically uncorrelated, which
is not perfectly applicable in the current case.  A compensating reduction in the effective degrees of freedom
from the nominal value of three would have the effect of marginally lowering the quoted CDF scale values relative to
the left-hand $\chi^2$ axis, slightly compressing the displayed error margins.

In the analysis presented in the second panel, we are instead intent on establishing a global lower bound on $M_{1/2}$ by
consideration of those studies without compelling hints of new physics.
The one-sided limit employed here is appropriate when the null hypothesis is disfavored only for deviations in a single direction, in this
case toward excessively large $\chi^2$ values.  The full exclusion is thus shifted toward a single tail of the
distribution, and the exclusion bounds are drawn directly at CDF values of $68\%$ and $95\%$.  Although matching bounds
beneath the median are also plotted for visual reference, they are less meaningful in this context.
To cope with the expected strong signal inter-dependence in this case, and to avoid
disproportionately emphasizing or suppressing the influence of either of the two broad search
strategies, we have opted to average the $\chi^2$ deviation at each $M_{1/2}$ across each closely related channel.
The results are then presented as a $\chi^2$ analysis in two effective or composite ``unit strength'' degrees of freedom.

The searches demonstrating an excess allow for the isolation of a best fit to the rate of \fsu5 SUSY production,
as regulated by the mass parameter $M_{1/2}$.  The five minima (which again hail from two broadly independent search strategies,
as analyzed in Table~\ref{tab:correlations}) are in excellent agreement, falling within the range of 756~GeV to 796~GeV.
The overall best fit for productive channels is at 792~GeV, and the intersections below and above with the median
Cumulative Distribution Function (CDF) probability span a range of about 730~GeV to 990~GeV.  This upper bound
is already beyond the range of $M_{1/2}$ values around 900~GeV at which the \fsu5 LSP becomes charged~\cite{Li:2011xu},
generating a well-defined finite truncation of the viable model search space.  The $2\sigma$ $\chi^2$ 
intersection for the inclusive searches sets a lower bound around 710~GeV.  The $2\sigma$ exclusion range from the non-productive channels is near
735~GeV in the second panel.  The depth of the $\chi^2$ well is rather remarkable, with the SM limit rising well above the median probability,
even without compensation for the reduction in effective degrees of freedom.

It is clear that this first round of data analyzed at the elevated collision energy
does seem to systematically isolate a somewhat heavier spectrum than that suggested by the
corresponding $\sqrt{s} = 7$~TeV~\cite{ATLAS-CONF-2012-033,ATLAS-CONF-2012-037} data,
previously analyzed in Refs.~\cite{Li:2012tr,Li:2012ix,Li:2012qv,Li:2012jf}.
This is identical to the statement that growth of signal relative to background is comparatively
suppressed in the higher energy results, and that the most optimistic expectations for a linear
extrapolation of the prior ratios have not in full materialized.
For comparison, a structurally similar study of the earlier LHC run~\cite{Li:2012ix} isolated a best
fit for $M_{1/2}$ at 658~GeV, with the median intersection stretching from 600~GeV to 770~GeV.
However, including the previously described error analysis, our parallel $\chi^2$ treatment of down-scaled ($\times 1/2$)
event production (not pictured) suggests a range for the intersection with the probability
median of the inclusive searches that extends from about 660~GeV to 930~GeV.  The $2\sigma$ intersections
for the inclusive and exclusive searches drop to approximately 620~GeV and 660~GeV, respectively.
We thus conservatively establish a lower bound for the \fsu5 gaugino mass of about $M_{1/2} \ge 660$~GeV,
corresponding to LSP, gluino and light stop masses of approximately 133~GeV, 890~GeV, and 725~GeV, respectively.
The 1/4 strength event down-scaling, which we judge to be a rather more extreme scenario, generates a best fit
at 651~GeV, intersections with the median CDF value around 560~GeV and 870~GeV, and intersections
with the $2\sigma$ line around 515~GeV and 560~GeV for the inclusive and exclusive searches, respectively.

Our survey of the $\sqrt{s}= 8$~TeV results thus suggest that the benchmark favored in the current report remains
marginally consistent with that isolated by the $\sqrt{s} = 7$~TeV data.  This statement is justified by the
existence of a satisfactory mutual overlap between the masses encapsulated by the median 50\% CDF intersections of
the two studies, without resorting to any event rescaling.  Moreover, the overall lower boundary extracted from
the 8~TeV data at a down-scaling of 1/2 is essentially identical to the nominally established best fit at 7~TeV.
The vital importance of individual model-specific comparisons to data is emphasized by a comparison of our derived
${\widetilde{\chi}_1^0}$, ${\widetilde{t}_1}$, and ${\widetilde{g}}$ \fsu5 mass limits with those established
under simplified SUSY model assumptions by the ATLAS collaboration.  The observed disparity suggests to us
that the simplified limits may indeed exaggerate the bounds ascribable to physically realistic models, particularly
with regards to the gluino mass, and that care should be exercised against their overly literal interpretation. 

If indeed the reported excess production above the background estimates is the result of new physics and
not attributed to background fluctuations, then the signal growth for larger statistics should remain roughly
proportional to the increase in luminosity. We project the future signal significance in Table~\ref{tab:103} and Table~\ref{tab:109} of
the five 8 TeV ATLAS 5.8 \fb searches exceeding the requirement
${\rm S = 2 \times (\sqrt{N_S + N_B} - \sqrt{N_B})} \ge 1 $. We choose the milestone
luminosities 10 \fbns, 15 \fbns, and 20 \fbns, all reachable
in 2012. Our method of projecting forward the ATLAS background, while serving our limited scope here
adequately, can only be as reliable as the expectation of statistical, dynamic and procedural stability
across the transition in luminosity and model. We also presume static ATLAS data cutting strategies in
Refs.~\cite{ATLAS-CONF-2012-103,ATLAS-CONF-2012-109} from 5.8 \fb to 20 \fbns. The
projections in Table~\ref{tab:103} and Table~\ref{tab:109} could be regarded as conservative, if we
take into account the sizeable downward fluctuation already witnessed in the data observations from 7 TeV
to 8 TeV, and we certainly do not consider an upside surprise in the signal significance going forward to be
improbable. For example, the ATLAS 7 TeV background for SUSY search 7j80 increased by a factor of 6.6 from 1
\fb~\cite{Aad:2011qa} to 4.7 \fb~\cite{ATLAS-CONF-2012-037}, and concurrently, the data
observations increased by a factor of 5.0, displaying reasonable consistency with the growth in
luminosity. By contrast, the ATLAS background increases by a factor of 3.3 from 5 \fb at 7 TeV to 5.8 \fb at
8 TeV, the same factor by which \fsu5 7j80 event counts increase in our simulations, but the data
observations only increase by a factor of 1.7~\cite{ATLAS-CONF-2012-103}. Hence, the rate of change of
the data observations from 1 \fb at 7 TeV to 5.8 \fb at 8 TeV has been markedly diminished with respect to the
corresponding rate of change of the background and luminosity over this same period.

Although higher energy collider operation does
induce larger SUSY cross sections and additional phase space headroom, the collected luminosity at 7 and 8~TeV
still remains comparable.
It remains to be seen whether these small discrepancies shall ultimately be attributed to primarily statistical or
systematic origins in either or both of the independently collected data sets. 
Accordingly, we await the LHC data forthcoming through the conclusion of 2012 that could clarify whether this dichotomy is
triggered by a downward fluctuation in the observed 8 TeV 5.8 \fb data or an upward fluctuation in the 7 TeV
observed data, possibly also convoluted with either an overestimated background for 8 TeV or
an underestimated background for 7 TeV.

\begin{table}[htp]
	\centering
	\caption{Projected signal significance using the metric ${\rm S = 2 \times (\sqrt{N_S + N_B} - \sqrt{N_B})}$ for 10
\fbns, 15 \fbns, and 20 \fb at $\sqrt{s} = 8$ TeV for the searches of Ref.~\cite{ATLAS-CONF-2012-103} that surpass the
condition ${\rm S = 2 \times (\sqrt{N_S + N_B} - \sqrt{N_B})} \ge 1$ for 5.8 \fb at 8 TeV, where ${\rm N_s}$ is the
number of signal events and ${\rm N_b}$ is the number of background events.}
		\begin{tabular}{|c||c|c|c||c|c|c||c|c|c|} \cline{2-10}
		\multicolumn{1}{c|}{} & \multicolumn{3}{|c||}{\rm 7j80} & \multicolumn{3}{|c||}{\rm 8j55} & \multicolumn{3}{|c|}{\rm 8j80}\\ \cline{1-10}
$	~{\rm fb^{-1}}~	$&$	{\rm ~N_s~}	$&$	{\rm ~N_b~	}$&$	{\rm ~~S~~	}$&$	{\rm ~N_s~	}$&${\rm 	~N_b~	}$&$	{\rm ~~S~~	}$&$	{\rm ~N_s~	}$&$	{\rm ~N_b~	}$&$	{\rm ~~S~~	}$ \\ \hline \hline	
$	10	$&$	5.5	$&$	3.1	$&$	\textbf{2.3}	$&$	6.1	$&$	5.9	$&$	\textbf{2.1}	$&$	1.5	$&$	0.2	$&$	\textbf{1.8}	$	\\ \hline
$	15	$&$	8.3	$&$	4.6	$&$	\textbf{2.9}	$&$	9.2	$&$	8.9	$&$	\textbf{2.6}	$&$	2.4	$&$	0.2	$&$	\textbf{2.3}	$	\\ \hline
$	20	$&$	11.0	$&$	6.2	$&$	\textbf{3.3}	$&$	12.3	$&$	11.8	$&$	\textbf{2.9}	$&$	3.1	$&$	0.3	$&$	\textbf{2.6}$	\\ \hline
		\end{tabular}
		\label{tab:103}
\end{table}

\begin{table}[htp]
	\centering
	\caption{Projected signal significance for the searches of Ref.~\cite{ATLAS-CONF-2012-109} that surpass the condition ${\rm S} \ge 1$ for 5.8 \fb at 8 TeV.}
		\begin{tabular}{|c||c|c|c||c|c|c|} \cline{2-7}
				\multicolumn{1}{c|}{} & \multicolumn{3}{|c||}{\rm SRE Loose} & \multicolumn{3}{|c|}{\rm SRE Medium} \\ \cline{1-7}
$	~{\rm fb^{-1}}~	$&$	{\rm ~N_s~}	$&$	{\rm ~N_b~	}$&$	{\rm ~~S~~	}$&$	{\rm ~N_s~	}$&${\rm 	~N_b~	}$&$	{\rm ~~S~~	}$ \\ \hline \hline	
$	10	$&$	5.7	$&$	9.8	$&$	\textbf{1.6}	$&$	6.0	$&$	6.0	$&$	\textbf{2.0}	$	\\ \hline
$	15	$&$	8.6	$&$	14.7	$&$	\textbf{2.0}	$&$	9.0	$&$	9.0	$&$	\textbf{2.5}	$	\\ \hline
$	20	$&$	11.3	$&$	19.7	$&$	\textbf{2.3}	$&$	12.0	$&$	12.0	$&$	\textbf{2.9}	$	\\ \hline
		\end{tabular}
		\label{tab:109}
\end{table}

\section{Primordial Synthesis}

We previously advertised a conspicuously correlated region of the \fsu5 model space adherent to
experimentally derived constraints imposed upon the set of rare-decay processes consisting of the
flavor changing neutral current processes $b \to s \gamma$ and \bs0, and the anomalous magnetic moment
$(g-2)_{\mu}$ of the muon, the intersection of which we labeled as the \textit{Golden Strip}. The recent
calculation of the complete tenth-order QED terms for $(g-2)_{\mu}$~\cite{Aoyama:2012wk} has
motivated a fresh inspection of these processes, expanding our original exploration for even deeper
correlative behavior within a wider scope of experiments presently searching for supersymmetry and both
direct and indirect evidence for dark matter. The natural synthesis of all these elements in an
\fsu5 framework exhibits clear global coherence.

The base of our investigation centers on the region of No-Scale \fsu5 model space satisfying a set of stable
bare-minimal experimental constraints~\cite{Li:2011xu}, defined by consistency with the
world-average top quark mass $172.2 \le m_t \le 174.4$ GeV~\cite{:1900yx}, No-Scale boundary
conditions $m_0 = A_0 = B_{\mu} = 0$, radiative electroweak symmetry breaking (EWSB), centrally observed
7-year WMAP cold dark matter density limits $0.1088 \le \Omega h^2 \le 0.1158$~\cite{Komatsu:2010fb},
and LEP constraints on the lightest CP-even Higgs boson, light SUSY chargino, and neutralino masses. The
union of all these bare-minimal constraints alone is certainly non-trivial, in particular the
convergence of the No-Scale boundary conditions, most notably the requirement of vanishing $B_{\mu}$,
with the precision empirically measured $1\sigma$ uncertainties on the top quark mass $m_t$ and relic
density $\Omega h^2$.

The LHC observations of an $m_h \sim 125$ GeV light Higgs
boson~\cite{:2012gk,:2012gu,Aaltonen:2012qt}, while celebrated as a significant milestone in
experimental particle physics, unexpectedly injected considerable turmoil into the landscape of
supersymmetric models. Whereas the MSSM contribution to the Higgs boson mass is woefully insufficient to
attain 125 GeV and concurrently generate a plausibly light and testable SUSY mass content at the 8 TeV LHC
(simultaneously avoiding the invocation of fine-tuning to once again escape the gauge-hierarchy problem) the very favorable (and
possibly necessary) contributions to the Higgs boson mass quite naturally supplied by the additional
vector-like flippon multiplets provide the desired boost in \fsu5 to produce an $m_h \sim 125$ GeV light
Higgs boson.
We have previously completed robust analyses of an $m_h \sim 125$ GeV Higgs boson in
\fsu5~\cite{Li:2011ab,Li:2012jf}, demarcating a narrowly contoured strip carved out of the larger
region formed by the application of the bare-minimal constraints. This strip of model space, clearly
delineated in Refs.~\cite{Li:2012yd,Li:2012qv,Li:2012jf}, comprises the 
intersection of an $m_h \sim 125$ GeV Higgs boson, $1\sigma$ variance on the WMAP 7-year relic density
measurements and world-average top quark mass, and adherence to the No-Scale high-energy boundary
conditions, the totality of which we shall heretofore refer to as the \textit{125 GeV Higgs Strip}.
Though the simple existence of such a strip is itself already noteworthy, a true model of nature must further
exhibit profound correlation between \textit{all} observable quantities, evident within the empirical
uncertainties to which measurements obtained from current state-of-the-art technology is bound.

We now seek to synthesize the 125 GeV Higgs Strip with the amalgamation of complementary supersymmetry
experiments, including our 8 TeV conclusions of the prior section. We begin with the original components
of our Golden Strip~\cite{Li:2010mi,Li:2011xu,Li:2011xg}, which are the key rare process limits on
\textit{Br}($b \to s \gamma$), \textit{Br}(\bs0), and $\Delta a_{\mu}$ on $(g-2)_{\mu}$ of the muon.
For $b \to s \gamma$, we use the latest world average of the Heavy Flavor Averaging Group (HFAG), BABAR,
Belle, and CLEO, which is $(3.55 \pm 0.24_{\rm exp} \pm 0.09_{\rm model}) \times
10^{-4}$~\cite{Barberio:2007cr}. An alternate approach to the average~\cite{Artuso:2009jw} yields
a slightly smaller central value, but also a lower error, suggesting
\textit{Br}$(b \to s\gamma) = (3.50 \pm 0.14_{\rm exp} \pm 0.10_{\rm model}) \times 10^{-4}$. See Ref.~\cite{Misiak:2010dz} for recent
discussion and analysis. The theoretical SM contribution at the next-to-next-to-leading order (NNLO)
is estimated at \textit{Br}$(b \to s \gamma) = (3.15 \pm 0.23) \times 10^{-4}$~\cite{Misiak:2006zs} and
\textit{Br}$(b \to s \gamma) = (2.98 \pm 0.26) \times 10^{-4}$~\cite{Becher:2006pu}. The addition of
these errors in quadrature provides the $2\sigma$ limits of
$2.86 \times 10^{-4} \le Br(b \to s \gamma) \le 4.24 \times 10^{-4}$. The recent precision improved LHCb constraints on the B-decay process \bs0 of
\textit{Br}(\bs0) $< 4.5(3.8) \times 10^{-9}$ at the 95\% (90\%) confidence level~\cite{Aaij:2012ac}
are employed here, though we find the entire viable \fsu5 parameter space lies comfortably below this
upper limit~\cite{Li:2012yd}. The new calculations of the tenth-order QED terms for the theoretical
prediction of $(g-2)_{\mu}$ engenders a favorable shift in $\Delta a_{\mu}$ in the context of \fsu5,
where we apply the $2\sigma$ uncertainty of $6.6 \times 10^{-10} \le \Delta a_{\mu} \le 41.4 \times
10^{-10}$. The $b \to s \gamma$ and $(g-2)_{\mu}$ effects reside at their lower boundaries in the 125 GeV
Higgs Strip, as they exert pressure in opposing directions on $M_{1/2}$ since the leading gaugino and
squark contributions to \textit{Br}($b \to s \gamma$) admit an opposite sign to the Standard Model term
and Higgs contribution. On the contrary, the effect is additive for the non-Standard Model contribution
to $\Delta a_{\mu}$, establishing an upper limit on $M_{1/2}$. The SUSY contribution to
\textit{Br}($b \to s \gamma$) cannot be excessively large such that the Standard Model effect becomes minimized, thus
necessitating a sufficiently large, or lower bounded, $M_{1/2}$.

\begin{figure*}[htp]
        \centering
        \includegraphics[width=0.77\textwidth]{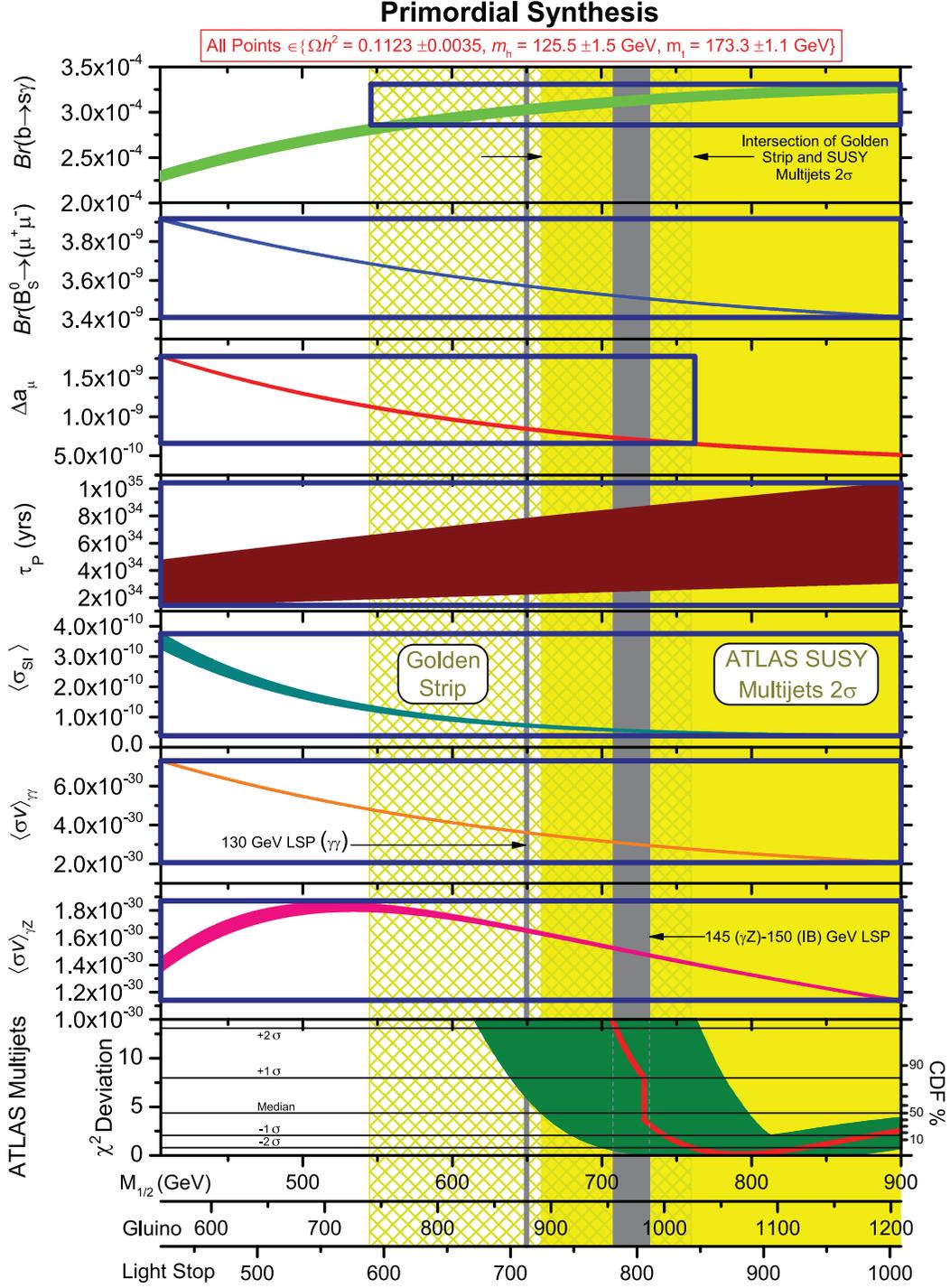}
        \caption{Primordial Synthesis of all currently progressing experiments searching for physics beyond the
Standard Model. All points depicted on each curve satisfy the conditions $0.1088 \le \Omega h^2 \le 0.1158$,
$124 \le m_h \le 127$ GeV, and $172.2 \le m_t \le 174.4$ GeV. Each curve thickness represents an uncertainty on
the strong coupling constant $0.1145 \le \alpha_s(M_Z) \le 0.1172$ (excluding the $\chi^2$ pane). The multi-axis
$\chi^2$ deviation in the bottom pane comprises an uncertainty derived from an increase and decrease by a factor
of 2 around the $\chi^2$ computed on the nominal number of \fsu5 events surviving all cuts (nominal value shown in
center of shaded curve). The cross-hatched region illustrates the Golden Strip, which identifies the intersection
of all current BSM experiments, excluding the ATLAS collider studies. The yellow region portrays the 2$\sigma$
uncertainty on the $\chi^2$ computed from \fsu5 Events $\times$ 0.5. We highlight the 130 GeV and 145-150 GeV LSP
masses by the vertical strips. Notice that we have inserted the gluino and light stop mass scales at the bottom
of the Figure to enable an instant conversion from the gaugino mass $M_{1/2}$, made possible by the
characteristic rescaling property of the \fsu5 SUSY spectrum in terms of $M_{1/2}$.}
        \label{fig:primordial}
\end{figure*}

The computation of the rare-decay processes for all points in the 125 GeV Higgs Strip are illustrated in
Figure (\ref{fig:primordial}). We implement a range on the strong coupling constant of $0.1145 \le \alpha_s(M_Z) \le 0.1172$
that tightly envelopes the central value of $\alpha_s(M_Z) = 0.1161$ that is
supported by recent direct observations~\cite{Bandurin:2011sh}, introducing a modest uncertainty
onto the calculation of each curve in Figure (\ref{fig:primordial}), represented by the contour
thickness in each pane. All SUSY particle masses, Higgs boson masses, relic densities, and constraints are computed with
{\tt MicrOMEGAs 2.4}, applying the proprietary modification of the {\tt SuSpect 2.34}
codebase to run the flippon-enhanced RGEs. In Figure (\ref{fig:primordial}), the boxed curve segments depict the experimentally observed
$2\sigma$ values.

We now expand our original Golden Strip to encompass proton decay and dark matter detection experiments.
The $p\rightarrow e^+ \pi^0$ mode in \fsu5 is depicted in Figure (\ref{fig:primordial}), indicative of
the large pervasive uncertainty propagated into the proton lifetime from the large QCD uncertainties in
$\alpha_s(M_Z)$. We apply the Super-Kamiokande established lower bound of $1.4 \times 10^{34}$ years at
the 90\% confidence level for the partial lifetime in the $p\rightarrow e^+ \pi^0$
mode~\cite{Hewett:2012ns}. For the spin-independent dark matter-nucleon cross section, the XENON100
experiment has probed down to $2\times 10^{-9}~{\rm pb}$ ($2\times 10^{-45}~{\rm cm^2}$) for a WIMP mass
of 55 GeV~\cite{Aprile:2012nq}, also at the 90\% confidence level. The No-Scale \fsu5 viable model space
shown in Figure (\ref{fig:primordial}) lies entirely below this upper bound~\cite{Li:2011in}.

Observations of a 130 GeV monochromatic gamma-ray line~\cite{Weniger:2012tx} by the FERMI-LAT Space
Telescope have stoked great interest into whether its origin results from dark matter annihilations
around the galactic center. The $\chi \chi \to \gamma \gamma$ line fits a WIMP mass $m_{\chi} \sim 130$
GeV~\cite{Weniger:2012tx}, while a $\chi \chi \to \gamma Z$ line fits a WIMP mass closer to
$m_{\chi} \sim 145$ GeV~\cite{Su:2012ft,Li:2012jf}. The fit is near $m_{\chi} \sim 150$ GeV for internal
bremsstrahlung~\cite{Bringmann:2012vr}. We further allow for the potential combination of all of the
above that could land a WIMP mass somewhere in the range $130 \lesssim m_{\chi} \lesssim 150$ GeV, and as a
consequence, we annotate the 130-150 GeV LSP mass region in Figure (\ref{fig:primordial}). The most
recent FERMI-LAT Collaboration upper bound on the gamma-ray annihilation cross section is
$\left\langle \sigma v \right\rangle \sim 10^{-26}~{\rm cm^3/s}$~\cite{Ackermann:2012qk}, which we
use in Figure (\ref{fig:primordial}), allowing for a possible subhalo boost factor, which can be on the
order of $\sim 1000$, as determined from examining extra-galactic clusters~\cite{Hektor:2012kc}.

We include in Figure (\ref{fig:primordial}) the multi-axis $\chi^2$ of the prior section of this work,
computed from those 8 TeV ATLAS multijet searches that display evidence of over-production above
background expectations. The vertical yellow band in Figure (\ref{fig:primordial}) depicts the
$2\sigma$ range around the $\chi^2$ minimum computed from the nominal number of \fsu5 simulated events
times 0.50, bordered by the lower $2\sigma$ boundary at about $M_{1/2} \sim 660$ GeV. The Golden Strip is
represented by the cross-hatched region, confined by the lower $2\sigma$ boundary on
\textit{Br}$(b \to s \gamma)$ at its lower $M_{1/2} \sim 545$ GeV limit, and by the lower $2\sigma$ boundary on
$\Delta a_{\mu}$ at the Golden Strip's upper $M_{1/2} \sim 760$ GeV limit. Demonstrated in Figure
(\ref{fig:primordial}) is the intersection of these two bands of model space defined by the
$2\sigma$ observable regions of completely uncorrelated experiments, though apparently exhibiting interesting
evidence of correlated behavior in a No-Scale \fsu5 framework. To further heighten the intrigue, the
130-150 GeV LSP model space corresponding to the FERMI-LAT Space Telescope observations of a 130 GeV
monochromatic gamma-ray line from the galactic center also very curiously lies snugly within the
intersection of all experiments. Notice that the gluino and light stop mass scales are inserted at the
bottom of Figure (\ref{fig:primordial}). Due to the characteristic rescaling property of No-Scale
\fsu5, a direct proportional relationship exists between the SUSY spectrum and gaugino $M_{1/2}$,
permitting a simple visual inspection of the associated gluino and light stop masses for any specified
$M_{1/2}$.

It is worth emphasizing again that all points delineated by the curves in each pane in Figure
(\ref{fig:primordial}) are themselves the intersection of three critical parameters measured to high
precision in current experiments, namely the 7-year WMAP relic density $0.1088 \le \Omega h^2 \le 0.1158$,
a 124-127 GeV light Higgs boson mass, and a $172.2 \le m_t \le 174.4$ GeV top quark mass. Therefore, at the
present time, we can find no experiment pertinent to the supersymmetric parameter space that is not in
conformance with the narrow band of No-Scale \fsu5 model space from $660 \lesssim M_{1/2} \lesssim 760$
GeV, which corresponds to sparticle masses of $133 \lesssim M({\rm \widetilde{\chi}_1^0}) \lesssim
160$ GeV, $725 \lesssim M({\rm \widetilde{t}_1}) \lesssim 845$ GeV, and
$890 \lesssim M({\rm \widetilde{g}}) \lesssim 1025$ GeV. Such a mutual interrelation between
all relevant experiments seems to strongly belie attribution to random stochastics.

The proximity of the 145-150 GeV LSP strip that resides within the theoretically and phenomenologically
favored \fsu5 parameter space defined by all model constraints, in relation to the minimum of our
multi-axis $\chi^2$ curve, recalls to mind a very similar level of statistical adjacency shared by the 
updated $\chi^2$ curves for the experimental Higgs boson mass measurements ($m_h \sim 125$ GeV) with 
the mass region theoretically and phenomenologically favored by electroweak precision measurements
at $m_h = 94^{+29}_{-24}$ GeV~\cite{LEPEWWG}.  The difference of about one standard deviation between the
empirically measured Higgs boson mass and the electroweak precision favored region is roughly akin to the
statistical margin separating the LHC SUSY multijet measurements and the optimum phenomenological \fsu5 region, where
we would assign, based upon the Figure~(\ref{fig:chi_square}) analysis, a standard fluctuation width of about 60~GeV to
deviations in the downward mass direction, and 200 GeV to the upper $\chi^2$ median intersection. Thus, we may take great satisfaction that such a level
of consistency is displayed between experiment and theory in \fsu5, supported by relevant historical precedent.

As two points of potentially relevant interest, we must also remark in passing on recent
developments regarding the measurement of the top quark mass and the strong coupling constant.
An external study based on ATLAS inclusive jet cross section data~\cite{Malaescu:2012iq} has suggested
the value $\alpha_s(M_Z) = 0.1151$, which is slightly lower than the world central value of 0.1161
on which we above remarked. Also, the CMS Collaboration has recently announced~\cite{:2012cz}
the world's single most precise top quark mass measurement at $m_h = 173.49 \pm 1.07$~GeV, with
a central value slightly above the existing world average. Moreover, the latest measurements by ATLAS show central values of $m_t = 174.5$
GeV~\cite{ATLAS:2012aj}, $m_t = 174.9$ GeV~\cite{ATLAS-CONF-2012-030}, and $m_t = 175.2$
GeV~\cite{ATLAS-CONF-2012-082}, all modestly elevated above the world average central value. In Ref.~\cite{Li:2012jf}, we investigated on the roles that a slightly elevated top quark mass, and a slightly reduced strong coupling could play in facilitating
satisfaction of the central Higgs mass measurements in the range of $125$--$126$~GeV, without
resorting to an overly heavy squark spectrum or extremities in the error margins for the Higgs mass itself.
The lowering of $\alpha_s$ while maintaining consistency with precision electroweak scale data
is an accommodation to which the flipped $SU(5)$ GUT is particularly well historically adapted~\cite{Ellis:1995at}.
An interesting side effect of this modification is an escalation in the proton decay rate linked 
to a parallel reduction in the GUT scale $M_{32}$.

We close our discussion of Figure (\ref{fig:primordial}) by remarking on the striking familiarity of
this figure to the correlation of predicted and observed light elemental abundances with the value of the
baryon-to-photon ratio given by the observations of the Cosmic Microwave Background (CMB) by WMAP. The
amazing consistency with which predictions of light element abundances by Primordial Nucleosynthesis
demonstrates with astronomical observations, while also compatible with the independently measured
CMB, provide powerful corroboration of the Big Bang Theory. We envision a compelling parallel here
amongst the synthesis of light elements predicted by Primordial Nucleosynthesis and observed by
experiments, with the synthesis in an ubiquitous \fsu5 structure in nature of all currently progressing
experiments searching for physics beyond the Standard Model, to which we aptly offer the description
\textit{Primordial Synthesis}. Analogous to the consistency encountered between theory and
experiment of light elemental abundances in Primordial Nucleosynthesis that provides a convincing
connection to the Big Bang Theory, we suggest that the consistency revealed in Figure
(\ref{fig:primordial}) between all the BSM experiments in No-Scale \fsu5 Primordial Synthesis
presents persuasive indications of BSM physics currently being probed at the LHC and indeed possibly all
the experiments involved in searching for the parameters in Figure (\ref{fig:primordial}).

\section{Conclusions}
We evaluated the first $\sqrt{s} = 8$ TeV 5.8 \fb ATLAS SUSY multijet data observations within the context
of No-Scale \fsu5 and suggested a simple mechanism for improving the efficiency of capture for SUSY multijet events.
The \fsu5 best SUSY mass spectrum fit to the 8 TeV data was found to be consistent with our 7 TeV results, within a
prescribed 2$\sigma$ margin of error.  These findings were synthesized with the complete
amalgamation of experiments currently searching for beyond the Standard Model physics, discovering
hidden correlations suggestive of deeper fundamental underpinnings. The Primordial Synthesis of all
BSM experiments uncovers a highly favorable region of the \fsu5 model space spanning from
$660 \lesssim M_{1/2} \lesssim 760$ GeV, corresponding to sparticle masses of
$133 \lesssim M({\rm \widetilde{\chi}_1^0}) \lesssim 160$ GeV, $725 \lesssim M({\rm \widetilde{t}_1}) \lesssim 845$ GeV,
and $890 \lesssim M({\rm \widetilde{g}}) \lesssim 1025$ GeV. We project that if indeed the production of
events beyond the Standard Model expectations in those active SUSY multijet searches studied here can be
attributed to new physics, then the completion of the 8 TeV run at the LHC in 2012 could provide
a strong indication of new physics.


\begin{acknowledgments}
We thank Tommaso Dorigo for helpful discussions on the LHC supersymmetry search. This research was
supported in part
by the DOE grant DE-FG03-95-Er-40917 (TL and DVN),
by the Natural Science Foundation of China
under grant numbers 10821504, 11075194, 11135003, and 11275246 (TL),
and by the Mitchell-Heep Chair in High Energy Physics (JAM).
We also thank Sam Houston State University
for providing high performance computing resources.
\end{acknowledgments}


\bibliography{bibliography}

\end{document}